\documentclass[twocolumn,showpacs,10pt]{revtex4-1}%
\usepackage{amsmath,amssymb,graphicx}
\usepackage{amsmath}
\usepackage{amsfonts}
\usepackage{amssymb}
\usepackage{graphicx}%
\providecommand{\U}[1]{\protect\rule{.1in}{.1in}}
\begin{document}
\title{True color night vision correlated imaging based on intensity correlation of light}

\author{Deyang Duan, Yunjie Xia}
\email{yjxia@qfnu.edu.cn}
\affiliation{School of Physics and Physical Engineering, Qufu Normal University, Qufu 273165, China\\
Shandong Provincial Key Laboratory of Laser Polarization and Information
Technology, Research Institute of Laser, Qufu Normal University, Qufu 273165, China}
\begin{abstract}
Night vision imaging is a technology that converts non-visible object to human
eyes into visible image in night and other low light environments. However,
the conventional night vision imaging can only directly produce grayscale
image. Here, we propose a novel night vision imaging method based on intensity
correlation of light. The object's information detected by infrared
non-visible light is expressed by visible light via the spatial intensity
correlation of light. With simple data processing, a color night vision image
can be directly produced by this approach without any pseudo-color image
processing. Theoretical and experimental results show that a color night
vision image comparable to classical visible light imaging quality can be
obtained by this method. Surprisingly, the color colorfulness index of the
reconstructed night vision image is significantly better than that of the
conventional visible light image and pseudo-color night vision image. Although
the reconstructed image can not completely restore the natural color of the
object, the color image obtained by this method is more natural sense than
that obtained by other pseudo-color image processing methods.
\end{abstract}
\maketitle
\section{Introduction}
Darkness is a non-visible barrier, which greatly limits the scope of human
vision. In order to broaden the scope of human vision, night vision imaging
has emerged. Night vision imaging is a kind of imaging technology, which can
transform the non-visible scene into visible image by using photoelectric
detection and imaging equipment under the condition of low light environments
[1]. Night vision imaging technology, which like turning night into day, has
greatly expanded human vision. Now, night vision imaging has been widely used
in military reconnaissance, security monitoring, automobile assisted driving
and other fields.

Since 1934, the first infrared image converter tube was invented by G. Holst
\emph{et.al}, night vision imaging technology has developed for nearly a
century. Because of its irreplaceable role in military affairs, night vision
imaging has been widely concerned since its birth. However, the conventional
night vision imaging technology still has some insurmountable difficulties,
e.g., the performance of infrared focal plane array (IRFPA) is quite lower
than that of charge-coupled device (CCD) and complementary metal oxide
semiconductor (CMOS); only grayscale image can be generated directly.
Consequently, the imaging quality of night vision imaging is far from that of
visible light imaging.

In this letter, we propose a novel night vision imaging method. In theory,
this method is based on the second-order intensity correlation of light field
[2-7]. In the experiment, one infrared laser and one visible laser are coupled
together to form a mixed laser beam that is modulated by a spatial light
modulator (SLM). Then, the modulated light is separated into two beams by a
dichroic mirror (DM). The infrared laser illuminates the object and the
reflected power is collected by a photomultiplier tube (PMT). The visible
laser does not interact with the object and is directly detected by a
conventional CCD camera. One monochromatic image is reconstructed by
cross-correlating the output signals of the two detectors. Thus, the object's
information detected by infrared non-visible light is expressed by visible
light via intensity correlation of light. Similarly, the other monochromatic
image can be produced by another pair of infrared laser and visible laser. A
colorful night vision image can be obtained via simple data processing. In
fact, the image of the object is produced by the CCD rather than the PMT
detector. Moreover, the IRFPA is not used in this method. Consequently, a
night vision image with high quality comparable to classical visible light
image can be produced by this method in theory.

Conventional night vision imaging techniques directly output grayscale images.
Then, the grayscale night vision image is transformed into monochrome or color
image by using image fusion processing to make it easier to observe [8-11]. In
conventional color night vision imaging, image fusion processing and visible
light reference image are essential [12-14]. However, a accurate visible light
reference image can not be obtained in some environments, which reduces the
effect of pseudo-color. In this approach, two monochromatic images can be
directly output via two pairs of infrared laser and a visible laser. These two
monochromatic images present different details and features of the object.
Through simple data processing, a color night vision image can be directly
produced without any pseudo-color image processing and visible light reference image.
\section{Theory}
We depict the method of true color night vision imaging based on intensity
correlation of light in Figure 1. Through the band-pass filters installed on
two filter wheels, the light illuminates on the SLM surface can be selected
arbitrarily. In this scheme, only one wavelength visible laser and one
wavelength infrared laser illuminate SLM at each time. For simplicity, we take
an infrared laser $E_{1}$ and a visible laser $E_{2}$ as examples to
illustrate the imaging process. The quasi-monochromatic infrared laser $E_{1}$
and the quasi-monochromatic visible laser $E_{2}$ are coupled together by DM3
to form a mixed beam. This mixed laser beam illuminates a SLM, and the
modulated light is separated into two beams by DM4. The object is illuminated
by an infrared laser, and the reflected light is detected by a PMT. The light
detected by the PMT can be expressed as%
\begin{align}
& E_{1}(x_{pmt},t)\nonumber\\
& =\int d\omega_{1}dq_{1}e^{-i\omega_{1}t}V(q_{1})E_{1}(\omega_{1}%
)H_{1}(x_{pmt},q_{1};\omega_{1})T(x_{o}).
\end{align}

The quantities $x$ and $q$ represent transverse position and wave vector,
respectively. The SLM produces spatial amplitude modulation of the light
represented by the random spatial mask function $V(q)$, which is taken to be
the same for $\omega_{1}$ and $\omega_{2}$ [15,16]. The functions $H_{1}$ is
transfer function that describe propagation through the DM4 to PMT. The
function $T(x)$ represents the object. The visible light reflected by DM4 does
not interact with the object and is directly detected by a conventional CCD
camera, which can be expressed as%
\begin{equation}
E_{2}(x_{ccd},t)=\int d\omega_{2}dq_{2}\,e^{-i\omega_{2}t}V(q_{2})E_{2}%
(\omega_{2})H_{2}(x_{ccd},q_{2};\omega_{2}),
\end{equation}
where the function $H_{2}$ describes propagation through the DM4 to CCD
camera. For later convenience we subtract the background, which comes from the
average intensities of light received by the PMT and the CCD by a DC block
[17,18]. \begin{figure}[ptbh]
\centering
\fbox{\includegraphics[width=1.0\linewidth]{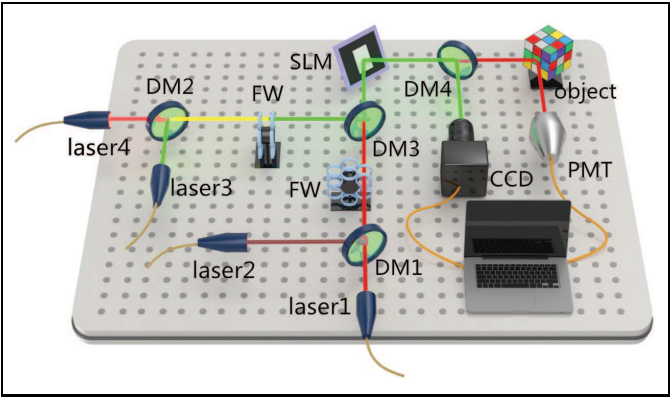}}\caption{The
diagrammatic sketch of true color night vision imaging based on intensity
correlation of light. DM: dichroic mirror, FW: filter wheel, SLM: spatial
light modulator, CCD: charge coupled device camera, PMT: photomultiplier
tube.}%
\label{fig:false-color}%
\end{figure}

The object's image is produced by measuring the intensity cross-correlation
function [19,20]. We thus obtain {\small
\begin{align}
&  G(x_{ccd},x_{pmt},t)\nonumber\\
&  =\left\langle \left\vert E_{1}(x_{pmt},t)\right\vert ^{2}\left\vert
E_{2}(x_{ccd},t)\right\vert ^{2}\right\rangle \nonumber\\
&  -\left\langle \left\vert E_{1}(x_{pmt},t)\right\vert ^{2}\right\rangle
\left\langle \left\vert E_{2}(x_{ccd},t)\right\vert ^{2}\right\rangle
\nonumber\\
&  =\int d\omega_{1}d\omega_{2}dq_{1}dq_{2}d\omega_{1}^{^{\prime}}d\omega
_{2}^{^{\prime}}dq_{1}^{^{\prime}}dq_{2}^{^{\prime}}\nonumber\\
&  \times H_{1}(x_{pmt},q_{1};\omega_{1})H_{1}^{\ast}(x_{pmt},q_{1};\omega
_{1})\\
&  \times H_{2}^{\ast}(x_{ccd},q_{2};\omega_{2})H_{2}(x_{ccd},q_{2};\omega
_{2})\nonumber\\
&  \times e^{-i(\omega_{1}-\omega_{1}^{^{\prime}})t}e^{-i(\omega_{2}%
-\omega_{2}^{^{\prime}})t}T(x_{o})T^{\ast}(x_{o})\nonumber\\
&  \times C(\omega_{1},\omega_{1}^{^{\prime}},\omega_{2,},\omega_{2}%
^{^{\prime}};q_{1},q_{1}^{^{\prime}},q_{2}{},q_{2}^{^{\prime}}{}),\nonumber
\end{align}
} where
\begin{align}
&  C(\omega_{1},\omega_{1}^{^{\prime}},\omega_{2,},\omega_{2}^{^{\prime}%
};q_{1},q_{1}^{^{\prime}},q_{2}{},q_{2}^{^{\prime}})\nonumber\\
&  =\left\langle E_{1}(\omega_{1})E_{1}(\omega_{1}^{^{\prime}})\right\rangle
\left\langle E_{2}(\omega_{2}^{^{\prime}})E_{2}(\omega_{2})\right\rangle \\
&  \times\left\langle V(q_{1})V(q_{1}^{^{\prime}})\right\rangle \left\langle
V(q_{2}^{^{\prime}})V(q_{2})\right\rangle \nonumber
\end{align}
is the intensity cross-correlations function in the spatial and temporal
frequency domain evaluated at the output surface of the SLM [15]. Here, the
mask function $V(q)$ is taken to possess spatial correlations that follow
Gaussian statistics [15,16]. Thus, the image expression can be rewritten as%
\begin{align}
&  G(x_{ccd},x_{pmt})\nonumber\\
&  =I_{1}I_{2}\int dx_{ccd}^{^{\prime}}dx_{pmt}^{^{\prime}}W(x_{ccd}%
^{^{\prime}},x_{pmt}^{^{\prime}})\\
&  \times H_{1}(x_{pmt},x_{pmt}^{^{\prime}};\omega_{1})H_{2}(x_{ccd}%
,x_{ccd}^{^{\prime}};\omega_{1})O(x_{o}),\nonumber
\end{align}
where $I_{a}=\left\langle \left\vert \int d\omega E_{a}\left(  \omega\right)
\right\vert ^{2}\right\rangle $ with $a=1,2$ being the product of the average
intensities of the detected light and the calculated light, respectively. The
function $W(x_{ccd}^{^{\prime}},x_{pmt}^{^{\prime}})$ is the spatial Fourier
transform of $\left\langle V(q_{pmt}^{^{\prime}})V^{\ast}(q_{ccd}^{^{\prime}%
})\right\rangle $. The transfer functions $H$ is written in position space.
$\left\langle T(x_{o})T^{\ast}(x_{o}^{^{\prime}})\right\rangle =\lambda
O(x_{o})\delta\left(  x_{o}-x_{o}^{^{\prime}}\right)  $.

Equation 5 shows that one night vision image is restructured via a pair of
infrared laser and visible laser. According to the color formation mechanism
of correlated imaging [16,21], this reconstructed night vision image show the
color of visible light rather than grayscale image produced by conventional
night vision imaging technology. Correspondingly, the other visible laser and
infrared laser can also produce a monochromatic night vision image. A color
night vision image can be obtained by processing these two monochromatic night
vision images. In this process, the indispensable pseudo-color image
processing in the conventional color night vision technology is not used in
this method.
\section{Experiment}
The experimental setup is illustrated in Figure 1. Two near-infrared lasers
with $\lambda_{1}=785$nm, $\lambda_{2}=830$nm (Changchun New Industries
Optoelectronics Technology Co., Ltd. MLL-III-785, MDL-III-830) are coupled
together to form a mixed beam by DM1 (Thorlabs, DMLP805). Then, this mixed
beam is filtered by two filers (Thorlabs FL05780-10, FL830-10) mounted on a
filter wheel (Daheng Xinjiyuan Technology Co., Ltd. GCM-14). Similarly, two
visible lasers with $\lambda_{3}=532$nm and $\lambda_{4}=635$nm (New
Industries MLL-III-532, MDL-III-635) are coupled together to form a mixed beam
by DM2 (Thorlabs, DMLP605). The output beam is filtered by two filers
(Huaxusheng NBP535, Thorlabs FL635-10) mounted on a filter wheel (Daheng
GCM-14). In this experiment, only one wavelength visible laser and one
wavelength infrared laser are illuminate on the surface of SLM at each time.
In this scheme, $\lambda_{1}=785$nm and $\lambda_{3}=532$nm as a pair of
beams, while $\lambda_{2}=830$nm and $\lambda_{4}=635$nm as a pair of beams
participate in imaging.  \begin{figure}[ptbh]
\centering
\fbox{\includegraphics[width=1.0\linewidth]{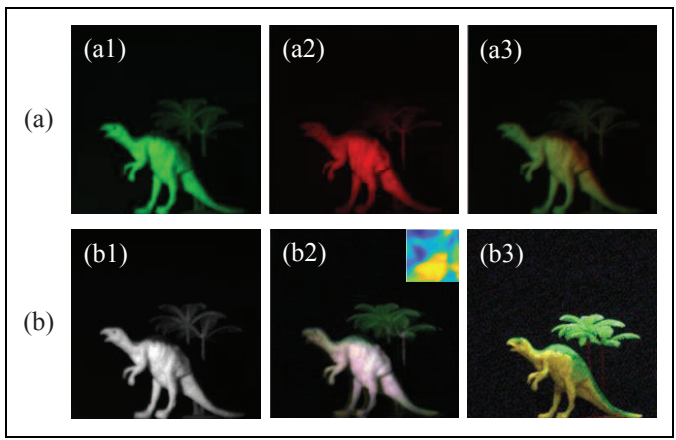}}\caption{top row: two
reconstructed monochromatic night vision images with different wavelengths,
(a1) $\lambda_{1}=785$nm and $\lambda_{3}=532$nm, (a2) $\lambda_{2}=830$nm and
$\lambda_{4}=635$nm, (a3) the reconstructed color night vision image. bottom
row: (b1) conventional night vision image, (b2) conventional color night
vision image produced by CbCr lookup table method (look-up table is shown in
the upper right corner), (b3) color image with visible light source.}%
\label{fig:false-color}%
\end{figure}In the following, we take the infrared laser $\lambda_{1}=785$nm
and the visible laser $\lambda_{3}=532$nm as an example to illustrate the
imaging process. The infrared laser and the visible laser are coupled together
by the DM3 (Huaxusheng LP690) to form a mixed beam. This mixed beam is
converged on the surface of two-dimensional amplitude-only ferroelectric
liquid crystal spatial light modulator (FLC-SLM, Meadowlark Optics
A512-450-850), with 512$\times$512 addressable 15$\mu m$ $\times$15$\mu m$
pixels. The wavelength modulation range of SLM is $450$nm-$850$nm. Then, the
modulated light is separated into two beams by the DM4 (Huaxusheng LP690 ).
The infrared laser beam illuminates an object and its reflected power is
detected by a PMT (Hamamatsu H10721-20). However, the visible light beam does
not interact with the object and is directly received by a conventional
visible light CCD camera (the imaging source DFK23U618). One green night
vision image is produced by cross-correlation the SLM signal and the output
signal of PMT. Correspondingly, we can obtain a red night vision image with
$\lambda_{2}=830$nm and $\lambda_{4}=635$nm.

Figure 2 compares the night vision image reconstructed by the novel method,
the conventional night vision image, and classical visible light image. Figure
2a1 and Figure 2a2 present the experimental result of two reconstructed
monochromatic images of different wavelengths with 200000 sets of data. Figure
2a3 is the color night vision image processed by the above two monochromatic
images. Figure 2b1 is a night vision image produced by conventional near
infrared camera (Intevac MicroVista-NIR). Figure 2b2 is a pseudo-color night
vision image generated by CbCr lookup table method [22]. Figure 2b3 is a
classical visible light image output by a CCD camera (the imaging source
DFK23U618). Figure 2a3 shows that color night vision image can be directly
produced by our approach. Figure 2a3 and Figure 2b2 obviously show the color
night vision image produced by our method does not completely restore the
natural color of the object, but the color night vision image obtained by this
scheme is more natural and friendly to human eyes than conventional software
pseudo-color method.

\begin{figure}[ptbh]
\centering
\fbox{\includegraphics[width=1.0\linewidth]{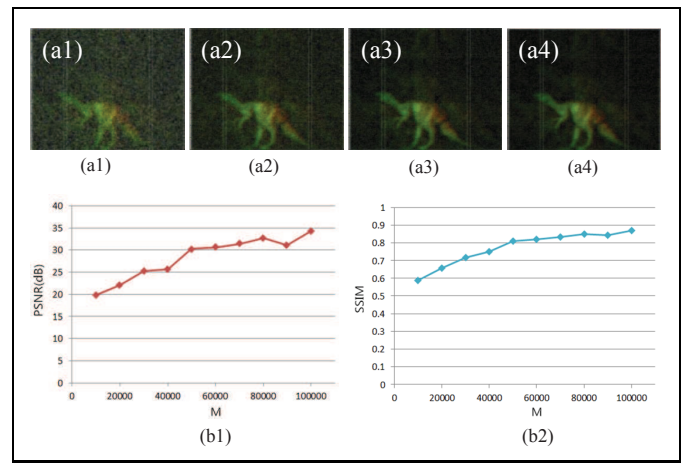}}\caption{Top row: The
reconstructed color night vision images with different realizations. The
numbers of frames are (a1) 10000, (a2) 40000, (a3) 70000, (a4) 100000. Bottom
row: The PSNR (b1) and SSIM (b2) curves of reconstructed image with different
realizations.}%
\label{fig:false-color}%
\end{figure}The image quality of this imaging method is significantly
dependent on the amount of data. Figures 3(a1-a4) present the reconstructed
color night vision images with different frames of data. To quantitatively
analyze the quality of the reconstructed image with different number of data,
peak signal to noise ratio (PSNR) and structural similarity index (SSIM) are
used as our evaluation index. Figure 3b shows that the image quality is
significantly improved by increasing the number of data. Certainly, the amount
of data used in reconstructing image can be greatly reduced by using image
processing and other methods [23-30]. \begin{figure}[ptbh]
\centering
\fbox{\includegraphics[width=0.8\linewidth]{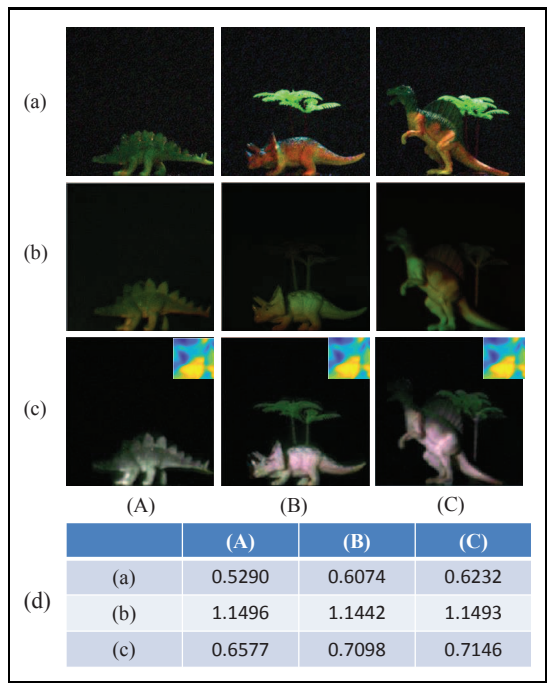}}\caption{Top row (a):
different objects (visible light image). Second row (b): The corresponding
reconstructed color night vision images produced by this method. Third row
(c): The corresponding color night vision images obtained by false color image
fusion processing. Bottom table: The CCI curves of the images with different
targets. }%
\label{fig:false-color}%
\end{figure}

Figure 4 compares the images obtained by visible light imaging, the novel
color night vision imaging, and conventional color night vision imaging based
on pasedu-color image fusion. Figures 4(a1- a3) show the three visible light
images. Figures 4(b1- b3) show that the three color images produced by this
night vision method. Figures 4(c1- c3) show the three color night vision
images obtained by pseud-color fusion processing with CbCr look-up method
[22]. The color colorfulness index (CCI) is used to quantitatively compare the
colorfulness of these three methods. It is surprising that the CCI of the new
color night vision imaging is significantly better than that of conventional
color night vision imaging and conventional color visible light imaging.
Figure 4(A-a) and Figure 4(A-b) show an interesting phenomenon that some
materials appear monochromatic in visible light, but the night vision images
obtained by this method can be color. Although the night vision images
produced by this method are only red, yellow and green to human eyes, as shown
in Fig. 4d, these different colors can be distinguished in computer vision.
\section{Summary}
In summary, true color night vision imaging based on intensity correlation of
light has been demonstrated in this article. Through the spatial intensity
correlation of light, the object's information detected by infrared
non-visible light is expressed by visible light. A color night vision image
comparable to classical optical imaging can be directly reconstructed without
any pseudo-color image processing and visible light reference image. Although
the reconstructed image is not completely restore the natural color of the
object, the color image obtained by this method is more natural and friendly
to human eyes than that obtained by other pseudo-color methods. Theoretically,
this scheme can obtain night vision images with the same quality as
conventional visible light imaging because the IRFPA is not used in this
scheme. A surprising found is the CCI of the reconstructed night vision image
is significantly better than that of conventional visible image, which seems
incredible because visible light image is generally considered to have the
richest colors. This new true color night vision imaging method provides a
promising solution to military reconnaissance, security monitoring and
automatic drive.
\section{Funding}
This project was supported by the National Natural Science Foundation of China
under Grant Nos. 11704221, 11574178 and 61675115, Taishan Scholar Project of
Shandong Province (China) under Grant No. tsqn201812059.

\end{document}